\documentclass[manuscript,screen]{acmart}
\AtBeginDocument{%
  \providecommand\BibTeX{{%
    \normalfont B\kern-0.5em{\scshape i\kern-0.25em b}\kern-0.8em\TeX}}}

\setcopyright{acmcopyright}
\copyrightyear{2022}
\acmYear{2022}
\acmDOI{XXXXXXX.XXXXXXX}

\acmConference[CHI ’22 Extended Abstracts]{In CHI Conference on Human Factors in Computing Systems Extended Abstracts (CHI ’22 Extended Abstracts)}{April 20 - May 05, 2022}{New Orleans, LA}
\definecolor{aziz}{RGB}{52, 76, 235}

\usepackage{caption}
\usepackage{subfigure}

\begin{document}

\title{Challenges and Opportunities of Content Optimization for Freeform User Interfaces}


\author{Aziz Niyazov}
\authornote{Both authors contributed equally to this research.}
\email{aziz.niyazov@irit.fr}
\affiliation{%
  \institution{IRIT - University of Toulouse}
  \country{France}
}

\author{Kaixing Zhao}
\authornotemark[1]
\email{kaixing.zhao@nwpu.edu.cn}
\affiliation{%
  \institution{School of Software, Northwestern Polytechnical University}
  \country{China}
}

\author{Tao Xu}
\email{xutao@nwpu.edu.cn}
\affiliation{%
  \institution{School of Software, Northwestern Polytechnical University}
  \country{China}
}

\author{Nicolas Mellado}
\email{nicolas.mellado@irit.fr}
\affiliation{%
  \institution{IRIT - CNRS}
  \country{France}
}

\author{Loic Barthe}
\email{loic.barthe@irit.fr}
\affiliation{%
  \institution{IRIT - University of Toulouse}
  \country{France}
}

\author{Marcos Serrano}
\email{marcos.serrano@irit.fr}
\affiliation{%
  \institution{IRIT - University of Toulouse}
  \country{France}
}

\renewcommand{\shortauthors}{Aziz Niyazov, et al.}

\begin{abstract}
While recent innovations on shape technologies allow for the creation of displays with almost unlimited form factors, current graphical user interfaces still rely on rectangular layouts and contents. This rectangular legacy hinders the progress of freeform displays, which are particularly relevant for pervasive scenarios to display interactive dynamic content where and when needed. By challenging the prevailing layout tradition on rectangular displays, freeform user interfaces raise design challenges which call for exploring the interlink between computational approaches and user interface generation and adaptation. In this position paper we report on previous work on content optimization for freeform user interfaces and anticipate the upcoming challenges and opportunities.
\end{abstract}



\begin{teaserfigure}
  \includegraphics[width=0.8\textwidth]{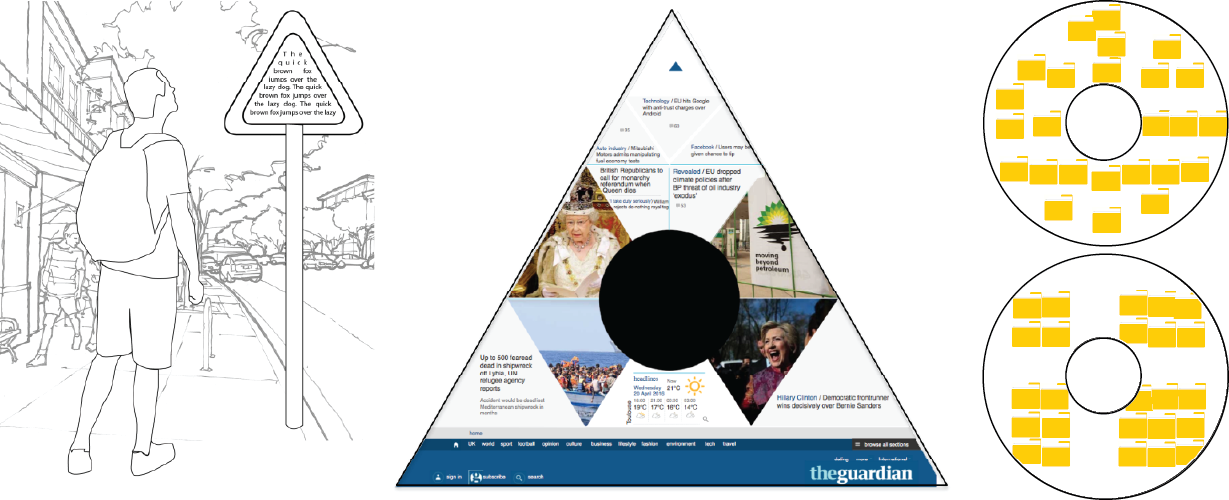}
  \caption{Three examples of previous projects on freeform interfaces. Left: a scenario illustrating the use of freeform interfaces to display text on road signs. \cite{Serrano:2017}; Middle: a designer-driven adaptation of a newspaper website to a triangular display \cite{Serrano:2016}; Right: Dynamic Decals allow to automatically position folder icons on non-rectangular displays using a constraint-based optimization approach \cite{Niyazov}}
  \Description{Three examples for freeform user interface.}
  \label{fig:teaser}
\end{teaserfigure}

\maketitle

\section{Introduction}
Graphical user interfaces are the most common mean to interact with computing devices, either be desktop PCs, smartphones, tablets or even wearable devices such as smartwatches. Most GUIs principles, such as WIMP interaction, have been defined decades ago and have little evolved. The introduction of mobile touchscreens could be seen as the last advancement that required designers and practitioners to rethink GUIs (for instance by adapting the size of icons to touch input). However, we are currently witnessing a revolution in display technologies, leading to commercially available curved (e.g. foldable phones) or freeform (e.g. circular smartwatches) displays. These novel displays can fit a variety of situations and are particularly relevant to fulfill the ubiquitous vision of displaying content everywhere \cite{10.5555/647987.741324}. Still, these developments involve heterogeneous display form factors which contradict the prevailing tradition of displaying content on rectangular GUIs.


Recent work in HCI has coined the term freeform user interface to refer to displays and interfaces having a non-rectangular form factor \cite{Niyazov, Serrano:2016, Serrano:2017}. The initial work on freeform interfaces mostly focused on the perceptual issues associated with displaying content on a non-rectangular display. For instance, authors conducted controlled studies to investigate text readibility \cite{Serrano:2016}, layout aesthetics \cite{Serrano:2017} or the impact on visual search \cite{Simon:2019}. After these initial explorations, which validated the fundamental design principles of how to display content on freeform interfaces, the looming question was how to adapt current practices to generate interactive and dynamic content, which are mostly based on toolkits designed for current rectangular UIs.


In this paper we report on a first exploration of the application of computational methods to dynamically optimize the placement of elements on a non-rectangular display. We then discuss future opportunities of freeform user interface based on our research experiences, and describe the remaining challenges, focusing on three important aspects of non-rectangular interfaces: GUI generation (organisation), adaptation and deformation. The goal of this position paper is to provide a phased summary to help researchers and practitioners address the research agenda on computational approaches for freeform interfaces.

\section{Past work: Dynamic Decals}



In this section, we give a brief presentation of previous work on generating and adapting freeform user interfaces. Our approach interactively optimizes the placements of graphical elements, so called decals, according to the interface geometry and desired layout properties, e.g. spacing or alignment. Each time the display property changes (e.g., the shape of the display changes, a physical object is moved on top of the display area), our system updates the decal positions.

Our contribution consisted in extending the use of \emph{arbitrary} constraints expressed as cost functions \cite{Mellado:2017}, which are minimized in the least squares sense to find a compromise between them. This enables the design of new constraints specific to GUI layout optimisation. Also, this approach allows to define a gamut constraint to force the decals to stay within the display area.

We first defined a set of desired GUI layout properties: GUI content visibility, layout simplicity (i.e. if items are aligned, the layout is simpler), and content grouping. 
Then we defined a set of new constraints to preserve these properties. We designed these constraints empirically, following the layout simplicity guidelines introduced by \citet{metrics, galitz}. In these constraints, coordinates and distances are expressed in pixels.

\begin{figure}[h]
\begin{center}
    \includegraphics[width=0.95\textwidth]{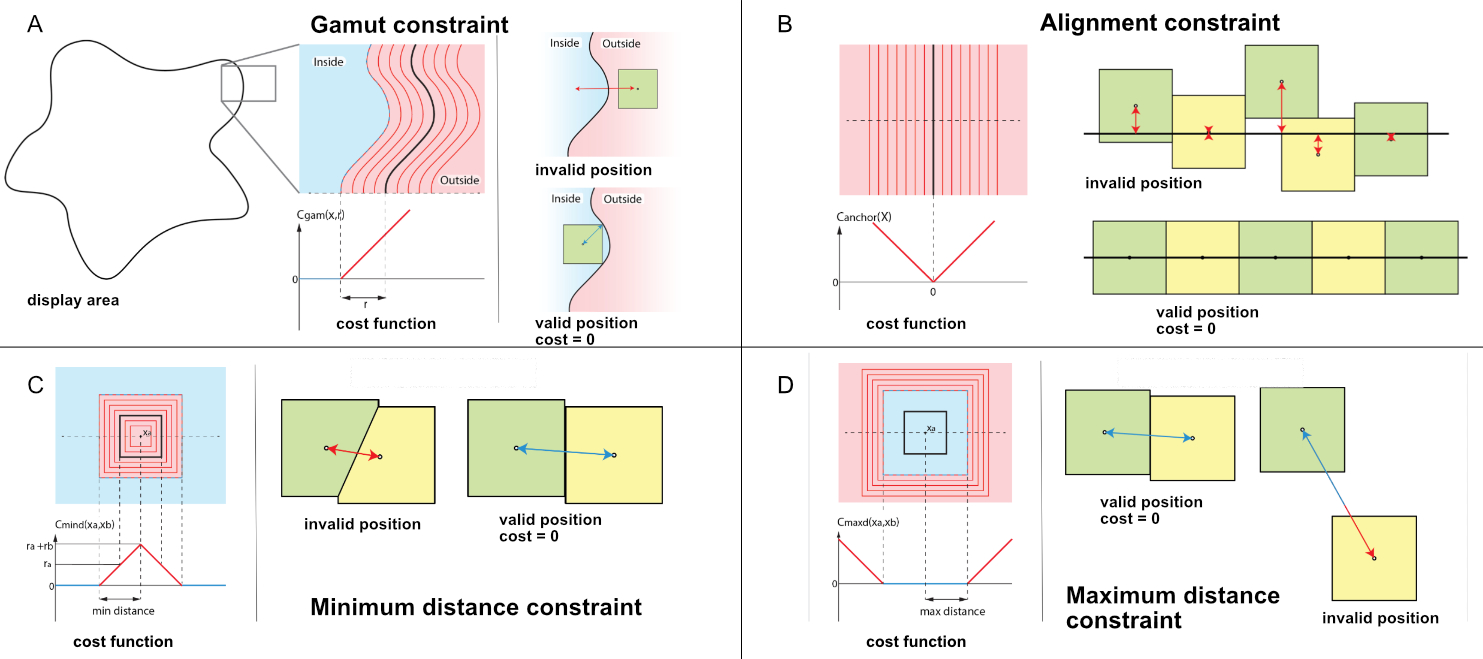}
\caption{Our layout constraints. For each one of them, we illustrate the cost function and the value of the cost according to the decals positions. }
\label{fig:layout-constraints}
\end{center}
\end{figure}

\emph{Gamut constraint}.
The goal of this constraint is to force the decals to stay within the display area (i.e. the gamut). A decal having a radius $r_i = n$ pixels is penalized if it lays \emph{inside} the display area but at less than $n$ pixels from the boundary (see Figure \ref{fig:layout-constraints}-A). In order to ensure fast evaluation with dynamic display area and decals with different radii, we compute a signed distance field $d_\mathcal{S}(\mathbf{x})$ to the display area boundary (negative inside, positive outside), and define the gamut constraint $c_\text{gam}(\textbf{x}, r)$ as follows:

\begin{equation}
c_\text{gam}(\textbf{x}, r) = 
\left\{ 
\begin{matrix}
  e_\text{step} + d_\mathcal{S}\left(\mathbf{x}\right) - r & \text{if } d_\mathcal{S}\left(\mathbf{x}\right) - r > 0 \\
  0  & \text{otherwise,}
\end{matrix}
\right.
\hspace{7mm} \text{and} \hspace{7mm}
d_\mathcal{S}(\mathbf{x}) = 
\left\{ 
\begin{matrix}
  & \left| \textbf{x} - \text{proj}_\mathcal{S}\left(\textbf{x}\right) \right|_2 & \text{if } \textbf{x} \notin \mathcal{S} \\
- & \left| \textbf{x} - \text{proj}_\mathcal{S}\left(\textbf{x}\right) \right|_2   & \text{otherwise.}
\end{matrix}
\right.
\end{equation}
The operator $\text{proj}_\mathcal{S}\left(\textbf{x}\right)$ projects $\textbf{x}$ on the boundary of the display area $\mathcal{S}$.
$e_\text{step}$ is a constant error term added to out-of-gamut positions, we used $e_\text{step}=10$ for all our experiments.

\emph{Distance constraints}.
We proposed two distance constraints in order to (1) prevent decals overlap to preserve content visibility and (2) favor decals grouping.
The \textbf{minimum distance} (see Figure \ref{fig:layout-constraints}-C) constraint penalizes overlapping decals. In order to ensure fast evaluation of this constraint, we assume without loss of generality that the decal's shape can be approximated by axis aligned square boxes of size $r_i$. Thus, overlap detection and cost can be efficiently computed using the $\text{L}_1$ distance between the decal's center:
\begin{equation}
    c_\text{mind}(\mathbf{x}_a, \mathbf{x}_b) = 
    min\left(0, \left|\mathbf{x}_a - \mathbf{x}_b\right|_1 - \left(r_a + r_b\right) \right) .
\end{equation}

The \textbf{maximum distance} (see Figure \ref{fig:layout-constraints}-D) constraint ensures that decals belonging to the same group remain in close proximity to each other. In contrast to $c_\text{mind}$, we look for proximity and do not want to favor distances in $x$ and $y$ directions (see Figure \ref{fig:layout-constraints}-D). 
Thus, we penalize decals when their L2 distance goes higher than a given threshold $d_\text{max}$:
\begin{equation}
    c_\text{maxd}(\mathbf{x}_a, \mathbf{x}_b) = 
    max\left(0, \left|\mathbf{x}_a - \mathbf{x}_b\right|_2 - d_\text{max} \right)  .
\end{equation}

\emph{Alignment}.
The anchor line constraint ensures that decals belonging to the same vertical/horizontal line remain aligned even when one of the decals moves (see Figure \ref{fig:layout-constraints}-B). The cost is computed w.r.t. the distance between the anchor line $l$ and each decal $\mathbf{x}_k$ belonging to a group denoted $\mathbf{X}$, as:
\begin{equation}
    c_\text{anchor}(\mathbf{X}) = \sum_k
    \left|\text{proj}_{l}\left(\mathbf{X}_k\right) - \mathbf{X}_k \right|_2 .
\end{equation}

Any group of decals can be associated with multiple lines, either vertical, horizontal or both (e.g. to form a grid).
Depending on the context, the position of the line can be either fixed (ie. coordinates are defined when creating the interface) or updated when the display properties change.

We validated our approach through an automatic evaluation that was followed and confirmed by a user study. In comparison to baselines (warping the entire interface or covering the content), the Dynamic Decals method improves interface aesthetics and content visibility while preserving content alignment and grouping.

\begin{figure}[h]
    \centering
    \includegraphics[width=0.90\textwidth]{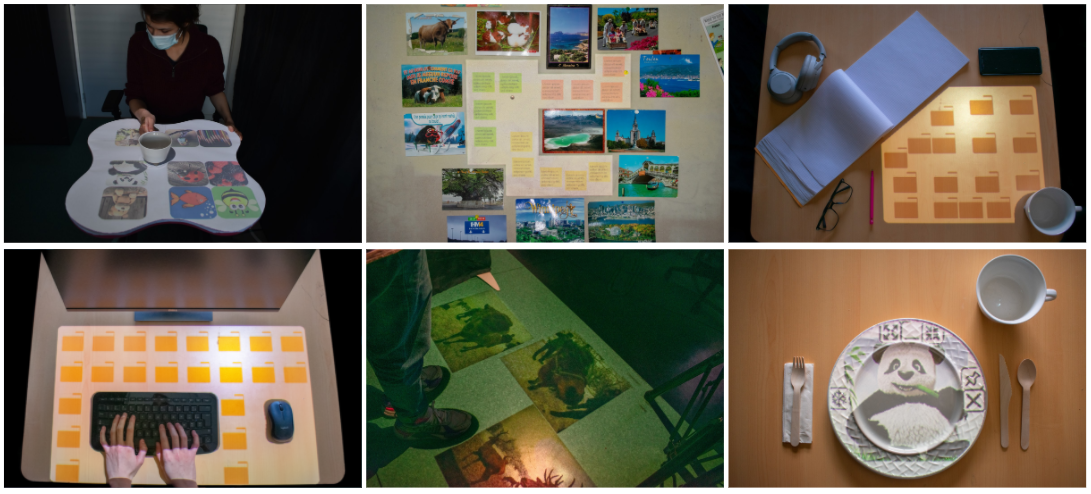}
    \caption{Illustrative application of the Dynamic Decals}
    \label{fig:applications}
\end{figure}

The application of Dynamic Decals might be vary various. Some examples are presented in Figure \ref{fig:applications} with different scenarios: interacting with a casual image gallery on freeform coffee table while drinking a cup of coffee, brainstorming using a wall projection combining virtual post-its with physical documents, personal augmented workspace that takes into account to the presence of physical objects, magic desk application \cite{Bi-2011} to extend the display area, floor projection of images that is aware of users position in a museum exhibit and an interface that adapts to a dishware.

\section{Towards Computational Approaches for Freeform User Interfaces}
We believe  that this prior work opens new promising directions for enriching HCI research with Computer Graphics methods and optimisation models for the design and implementation of new freeform GUIs. In this section, we discuss a set of potential challenges and opportunities for the generation and optimization of freeform user interfaces.

\subsection{Challenges for Freeform User Interface Generation and Optimisation}

\begin{itemize}
  
  \item \textit{Lack of fundamental knowledge on layout properties}. Unlike traditional rectangular interfaces layouts who have been designed and studied for decades, research on freeform user interfaces has only begun to receive attention in recent years. Therefore, there is still a lack of fundamental knowledge in this domain, especially on freeform layout properties. Although preliminary work has explored the impact of displaying text  \cite{Serrano:2016} and laying out content \cite{Serrano:2017} on freeform interfaces, these works only explored a limited set of display shapes. The number of possible display shapes is almost unlimited, specially when considering projection based systems where the overall display shape is defined by the available display surface and the presence of physical objects occluding the display area. 

 \item \textit{Generating interfaces for a large variety of shapes}. Traditional rectangular user interfaces follow a set of unified design criteria, e.g. WIMP interfaces. However, current freeform user interfaces scenarios illustrate a large variety of interface and display shapes. This makes it very difficult to derive a unified set of design requirements. Computational approaches that will be able to infer and generalize design properties from one display shape to another represent the most viable solution to this problem.

 \item \textit{Aesthetics metrics}. Aesthetics or visually pleasing compositions of graphical user interfaces play a crucial role in content perception and interaction. The aesthetic metrics are well defined for rectangular screens \cite{metrics}. However, changing the overall display shape and consequently adapting the content has been shown to have an impact on perceived aesthetics. There is still a lack of knowledge on how existing interface metrics formulas can be adapted to freeform interfaces, or if there will be a need to introduce new metrics. While some metrics such as density can directly apply to freeform displays, others such as balance, symmetry or simplicity might be perceived differently on such non-rectangular interfaces. To develop computational approaches that preserve the aesthetic properties, there is a need to define aesthetic metrics. 

 \item \textit{GUI implementation challenges}. Contrary to traditional rectangular interfaces, which can be developed using well established graphical toolkits, the development of freeform user interfaces is challenging due to the lack of well adapted methods. For the Dynamic Decals project we worked with computer graphic experts to redefine the notion of graphical icon at the pixel level. Instead of simply using predefined graphical items (e.g. buttons, widgets, panels), UI designers and developers need to be able to understand and develop novel methods to redefine the graphical aspect of interfaces at various levels of granularity. However current UI developers usually have no expertise in computer graphics. The development of the freeform interfaces will require profound changes in the training of future UI developers.

\end{itemize}

\subsection{Opportunities Emerging from the Development of Freeform User Interfaces}
The developments conducted to solve the various challenges faced by both academia and industry on freeform user interfaces can lead to a set of new opportunities for the HCI community.

\begin{itemize}
  
\item \textit{Generalize content-optimization for HCI}. The specific needs of freeform user interfaces call for adopting computational approaches such as dynamic layout adaptation using constraint-based optimization. The research conducted within the field of freeform interfaces could later be generalized to other types of displays or interfaces. For instance immersive visualizations can be seen as a particular instance of freeform displays, where the visualisation can be placed anywhere in the 3D environment. Our approaches to optimize content on 2D freeform interfaces could be extended to such 3D environments to properly layout content around the user while avoiding occluding relevant real worlds objects for instance.

\item \textit{Expand our knowledge on interactive optimization}.
Designing interfaces that adapt to contextual factors, such as the ongoing task, environment, or user perception, is a very complex task. In current approaches, content is usually assigned to a fixed position following an optimization approach and further manually adjusted by the users. Giving the opportunity to the user to adjust the optimisation itself will make content optimization more flexible, and can be extended to other interactive applications.

\item \textit{Develop interdisciplinary research}. As said before, developing freeform user interfaces is complex and often requires multi-disciplinary knowledge. As such, this field represents a very interesting opportunity for conducting interdisciplinary research and bringing together different communities beyond HCI. First, Computer Graphics methods are necessary to redefine the graphical aspect of these UIs. Second, cognitive science methods are need to understand the cognitive implications (information acquisition, visual search, spatial memory) of freeform user interfaces. Finally, Computational Approaches such as ML or constraint-based optimization are needed to generate and adapt freeform UIs.

\end{itemize}

\section{Conclusion}
In this paper, we discussed the challenges and future opportunities of freeform user interface based on our previous research experiences in this domain. The research challenges are the lack of fundamental knowledge on layout properties, the large variety of shapes, the lack of aesthetics metrics and the complexity of implementing freeform GUIs. We discussed two emerging opportunities resulting from addressing the previous challenges. First, we believe the proposed solutions could adapt to other fields of HCI such as immersive interfaces. Second, we believe work on freeform UIs represents a very interesting opportunity to develop interdisciplinary research.  We hope by discussing the challenges and opportunities of freeform user interface, we can inspire more researchers to contribute to this research domain and also promote its industrial development.

\section{Acknowledgments}

This work was supported by the ANR JCJC PERFIN grant (ANR-18-CE33-0009). We  thank  the  reviewers  of  this article for their relevant comments and recommendations. 


\bibliographystyle{ACM-Reference-Format}
\bibliography{sample-base}


\end{document}